\documentclass[twocolumn,notitlepage,nofootinbib]{revtex4-2}
\usepackage{amsmath,amssymb,bm,graphicx,hyperref,slashed}
\allowdisplaybreaks[1]

\renewcommand\d{\partial}
\newcommand\grad{\bm{\nabla}}
\newcommand\eps{\varepsilon}
\newcommand\0{\bm{0}}
\newcommand\A{\bm{A}}
\newcommand\B{\bm{B}}
\newcommand\E{\bm{E}}
\renewcommand\b{\bm{b}}
\renewcommand\j{\bm{j}}
\renewcommand\k{{\bm{k}}}
\newcommand\p{{\bm{p}}}
\renewcommand\r{\bm{r}}
\renewcommand\v{\bm{v}}
\newcommand\Z{\mathbb{Z}}
\newcommand\C{\mathcal{C}}
\renewcommand\P{\mathcal{P}}
\newcommand\reg{\mathrm{reg}}
\DeclareMathOperator\diag{diag}
\DeclareMathOperator\tr{tr}

\begin{document}

\title{Dynamical chiral magnetic current and instability in Weyl semimetals}

\author{Tatsuya Amitani}
\author{Yusuke Nishida}
\affiliation{Department of Physics, Tokyo Institute of Technology,
Ookayama, Meguro, Tokyo 152-8551, Japan}

\date{July 2022}

\begin{abstract}
Weyl semimetals realize massless relativistic fermions with two Weyl nodes separated in energy and momentum space, whose low-energy physics is described by Dirac fermions with an axial gauge constant.
Here, we study their electromagnetic linear responses based on the effective field theory and on the chiral kinetic theory.
Although the static chiral magnetic effect is canceled by the Chern-Simons current under the Pauli-Villars regularization, a dynamical magnetic field is found capable of driving an electric current along its direction, with the total transported charge being independent of temperature and chemical potential for a uniform field.
We also incorporate dissipation in the relaxation-time approximation and study collective excitations coupled with Maxwell electromagnetic fields when Weyl node populations deviate from equilibrium.
Their dispersion relations at low frequency and long wavelength are determined only by electric, chiral magnetic, and anomalous Hall conductivities, which predict unstable modes leading to anisotropic generation of electromagnetic waves oriented to the direction of Weyl node separation.
\end{abstract}

\maketitle
%\tableofcontents

\section{Introduction}
Dirac and Weyl semimetals constitute a new class of topological materials in three dimensions, where the valence and conduction bands touch each other at isolated points in the Brillouin zone protected by topology and symmetry~\cite{Armitage:2018,Gorbar-Miransky-Shovkovy-Sukhachov}.
The dispersion relation in the vicinity of one band touching point is generally linear in momentum so as to realize massless relativistic fermions.
Because all states under time-reversal and inversion symmetries are doubly degenerate due to the Kramers theorem, each band touching point is effectively described by the Dirac equation~\cite{Young:2012,Wang:2012}.
However, when such symmetries are broken, the band touching point is to be separated into a pair of nondegenerate Weyl nodes with opposite chiralities~\cite{Nielsen:1981a,Nielsen:1981b}, which are now described by the right- and left-handed Weyl equations~\cite{Murakami:2007,Wan:2011}.
Consequently, Dirac and Weyl semimetals offer exciting platforms to study relativistic dynamics of chiral fermions in condensed matter systems.

Among others, the chiral magnetic effect has attracted broad interest across diverse fields in physics~\cite{Miransky:2015}, which predicts the electric current along an external magnetic field in the presence of an axial chemical potential~\cite{Kharzeev:2008,Fukushima:2008}:
\begin{align}\label{eq:CME}
\j_\mathrm{CME} = \frac{e^2}{2\pi^2}\mu_5\B.
\end{align}
This current is nondissipative if it exists in equilibrium, being one celebrated example of anomalous transport phenomena in relativistic quantum matters sourced from the axial anomaly~\cite{Vilenkin:1980,Nielsen:1983}.
However, after resolving some confusion, it is now established that the chiral magnetic effect is absent in equilibrium~\cite{Vazifeh:2013,Basar:2014,Landsteiner:2016}.
Therefore, in order to activate the chiral magnetic effect, one needs to drive the system out of equilibrium.
For example, by applying an external electric field in Dirac and Weyl semimetals~\cite{Son:2013a,Burkov:2014}, possible signatures of the chiral magnetic effect have been experimentally observed~\cite{Armitage:2018,Gorbar-Miransky-Shovkovy-Sukhachov}.
It was also proposed to induce the chiral magnetic effect transiently by distorting a crystal with strain~\cite{Cortijo:2016,Pikulin:2016} or by heating Weyl nodes unequally under a strong electric field~\cite{Nandy:2020}.

What we explore in this work is another possible form of the chiral magnetic current in Weyl semimetals that is driven by a dynamical magnetic field.
In order to describe their low-energy physics, study electromagnetic linear responses, and derive physical consequences, both the effective field theory~\cite{Grushin:2012,Zyuzin:2012,Goswami:2013} (Sec.~\ref{sec:current}) and the chiral kinetic theory~\cite{Stephanov:2012,Son:2012,Son:2013b,Gorbar:2017a} (Sec.~\ref{sec:instability}) will be complementarily employed.
We set $\hbar=k_B=1$ throughout this paper and the Minkowski metric is chosen to be $\eta_{\mu\nu}=\diag(1,-1,-1,-1)$.
Greek indices such as $\mu,\nu,\dots$ are valued at $0,1,2,3$, whereas Latin $i,j,\dots$ are at $1,2,3$, and repeated indices are implicitly summed.
An integration over three-dimensional momentum $\p$ is denoted by $\int_\p\equiv\int d^3\p/(2\pi)^3$ for the sake of brevity.

\begin{figure}[b]
\includegraphics[width=0.85\columnwidth]{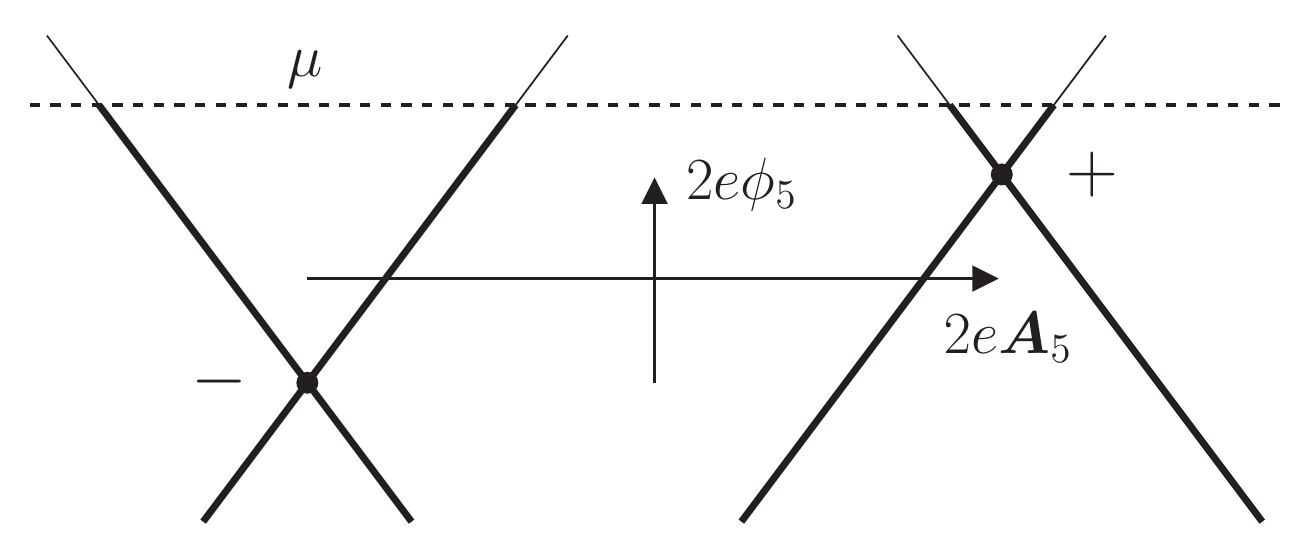}
\caption{\label{fig:nodes}
Two Weyl nodes with chirality $\chi=\pm$ separated in energy (vertical) and momentum (horizontal) directions by $2e\phi_5$ and $2e\A_5$, respectively.
The axial chemical potential is fixed at $\mu_5=-e\phi_5$ in equilibrium.}
\end{figure}

\section{Chiral magnetic current}\label{sec:current}
Low-energy effective description of Weyl semimetals is provided by massless Dirac fermions whose field-theoretical action reads
\begin{align}
S = \int d^4x\,\bar\psi(x)
\bigl[i\slashed\d - e\slashed A_5\gamma^5 - e\slashed A(x)\bigr]\psi(x).
\end{align}
Here, $e=-|e|$ is the electron charge, $A^\mu(x)$ is an electromagnetic gauge field, and an axial gauge constant $A_5^\mu$ amounts to introducing a separation between two Weyl nodes as shown in Fig.~\ref{fig:nodes}~\cite{Chernodub:2022}.
We note that $v$ entering $x^0=vt$, $A^0(x)=\phi(x)/v$, and $A_5^0=\phi_5/v$ is the Fermi velocity instead of the speed of light.

\begin{figure}[t]
\includegraphics[width=0.9\columnwidth]{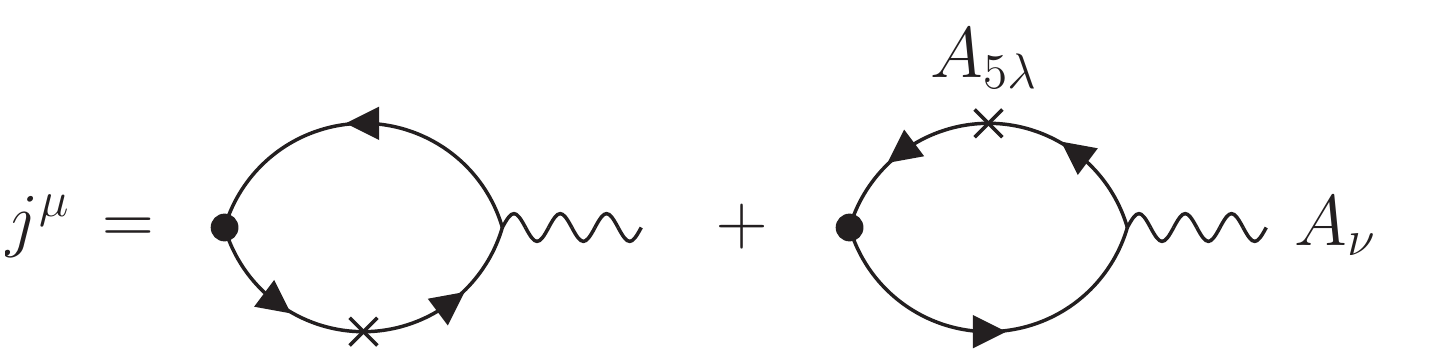}
\caption{\label{fig:current}
Feynman diagrams contributing to the current density in Eq.~(\ref{eq:current}).}
\end{figure}

In the linear response theory with respect to $A^\nu(x)$, the current density at the linear order in $A_5^\lambda$ is provided by
\begin{align}\label{eq:current}
\tilde j^\mu(k) = e^3\,\Pi^{\mu\nu\lambda}(k)A_{5\lambda}\tilde A_\nu(k),
\end{align}
where $\Pi^{\mu\nu\lambda}(k)$ is the retarded correlation function depicted in Fig.~\ref{fig:current} and the tildes indicate the space-time Fourier transforms.
The retarded correlation function is most conveniently obtained from the corresponding imaginary-time-ordered correlation function~\cite{Altland-Simons},
\begin{align}
& \Pi^{\mu\nu\lambda}(k) \notag\\
&= T\sum_{p_0}\int_\p\tr\!\left[\frac1{\slashed p+\slashed k-m}\gamma^\nu
\frac1{\slashed p-m}\gamma^\lambda\gamma^5\frac1{\slashed p-m}\gamma^\mu\right]_\reg \notag\\
&\quad + T\sum_{p_0}\int_\p\tr\!\left[\frac1{\slashed p-m}\gamma^\lambda\gamma^5
\frac1{\slashed p-m}\gamma^\nu\frac1{\slashed p-\slashed k-m}\gamma^\mu\right]_\reg,
\end{align}
where $vk_0=i2n\pi T$ and $vp_0=i(2n+1)\pi T+\mu$ with $n\in\Z$ are the bosonic and fermionic Matsubara frequencies, respectively, and the former is to be analytically continued into $vk_0\to\omega^+\equiv\omega+i0^+$.
In order to define the superficially divergent integral, we employ the Pauli-Villars regularization, which indicated by the subscript ``reg'' replaces the integrand as
\begin{align}
\int\,[h(m)]_\reg \equiv \lim_{m\to\infty}\int\,[h(0)-h(m)],
\end{align}
with the limit taken after the integration~\cite{Bertlmann}.
This regularization amounts to subtracting the contribution of massive Pauli-Villars ghosts from that of massless Dirac fermions.
Although the current density at the zeroth order in $A_5^\lambda$ is not presented here, we confirmed that it has the same structure as found in Eq.~(\ref{eq:kinetic}) below.

We then evaluate the trace of gamma matrices, the Matsubara frequency summation, the angular integration of $\p$, and its radial integration in part, along the line of analysis described in Ref.~\cite{Amitani:2022}.
After the lengthy but straightforward calculations, we find
\begin{align}\label{eq:response}
\tilde j^\mu(k) &= -\frac{e^3v}{2\pi^2}\epsilon^{\mu\lambda\kappa\nu}
A_{5\lambda}ik_\kappa\tilde A_\nu(k) \notag\\
&\quad + \frac{e^3}{2\pi^2}\int_0^\infty\!d\eps\,N'_+(\eps)
f_\eps(\omega^+\!,|\k|)\delta^\mu{}_i\phi_5\tilde B^i(k),
\end{align}
where $\epsilon^{\mu\lambda\kappa\nu}$ for $\epsilon^{0123}=1$ is the totally antisymmetric tensor, $\delta^\mu{}_i$ is the Kronecker delta, $N_+(\eps)\equiv n_T(\eps-\mu)+n_T(\eps+\mu)$ as well as $N'_+(\eps)=dN_+(\eps)/d\eps$ is introduced with $n_T(\eps)=1/(e^{\eps/T}+1)$ being the Fermi-Dirac distribution function, and
\begin{align}
f_\eps(\omega,k) &\equiv \frac{\omega^2-(vk)^2}{4(vk)^2}
\left[\frac{\omega}{vk}\ln\!\left(\frac{(\omega+vk)^2[(\omega-vk)^2-4\eps^2]}
{(\omega-vk)^2[(\omega+vk)^2-4\eps^2]}\right)\right. \notag\\
&\quad\left.{} + \frac{2\eps}{vk}
\ln\!\left(\frac{\omega^2-(vk-2\eps)^2}{\omega^2-(vk+2\eps)^2}\right)\right].
\end{align}
Whereas massless Dirac fermions contribute to both terms on the right-hand side of Eq.~(\ref{eq:response}), massive Pauli-Villars ghosts contribute to the first term only.
The first term is the so-called Bardeen-Zumino or Chern-Simons current~\cite{Landsteiner:2014,Landsteiner:2016}, which provides
\begin{subequations}\label{eq:chern-simons}
\begin{align}
\delta\rho(x) &= \frac{e^3}{2\pi^2}\A_5\cdot\B(x), \\
\delta\j(x) &= -\frac{e^3}{2\pi^2}\A_5\times\E(x) + \frac{e^3}{2\pi^2}\phi_5\B(x)
\end{align}
\end{subequations}
for the charge and current densities.
Its correct form was unavailable without the regularization, which is consistent with the observation in a lattice model of Weyl semimetals~\cite{Gorbar:2017b}.
On the other hand, the second term accompanied by the distribution functions is found to provide the current density $\sim\phi_5\B(x)$ only, which is nontrivial because other terms such as $\A_5\times\E(x)$ and $(\A_5\cdot\grad)\dot\B(x)$ are allowed from the symmetry perspective.

In order to shed light on the possible current density along the magnetic field, we first consider the static followed by uniform limit,
\begin{align}
\lim_{k\to0}\lim_{\omega\to0}f_\eps(\omega,k)
= \lim_{k\to0}\frac\eps{2vk}\ln\!\left(\frac{(vk+2\eps)^2}{(vk-2\eps)^2}\right)
= 1,
\end{align}
which leads to
\begin{align}
\j|_\text{static} = -\frac{e^3}{2\pi^2}\A_5\times\E.
\end{align}
Therefore, there remains only the anomalous Hall effect~\cite{Yang:2011,Burkov:2011} and the chiral magnetic effect as of Eq.~(\ref{eq:CME}) is absent in equilibrium~\cite{Vazifeh:2013,Basar:2014,Landsteiner:2016}, for which the regularization canceling the second term in Eq.~(\ref{eq:response}) is essential.
On the other hand, when the uniform followed by static limit,
\begin{align}
\lim_{\omega\to0}\lim_{k\to0}f_\eps(\omega,k)
= \lim_{\omega\to0}\frac{4\eps^2(4\eps^2-3\omega^2)}{3(4\eps^2-\omega^2)^2}
= \frac13,
\end{align}
is considered, we obtain
\begin{align}\label{eq:uniform}
\j|_\text{uniform} = -\frac{e^3}{2\pi^2}\A_5\times\E + \frac{e^3}{3\pi^2}\phi_5\B.
\end{align}
The chiral magnetic current is now present~\cite{Chen:2013}, although its coefficient is smaller in magnitude by $2/3$ and opposite in sign compared to Eq.~(\ref{eq:CME}) for $\mu_5=-e\phi_5$.
The resulting current is, so to say, the dc limit of ``dynamical chiral magnetic current,'' which is possible by driving the system out of equilibrium with a dynamical magnetic field as well as with an induced electric field.
This situation should be contrasted with the chiral magnetic effect in Eq.~(\ref{eq:CME}), which assumes a static magnetic field applied to nonequilibrium states with $\mu_5\neq0$ under $e\phi_5=0$~\cite{Kharzeev:2009,Landsteiner:2016}.

\begin{figure}[t]
\includegraphics[width=0.9\columnwidth]{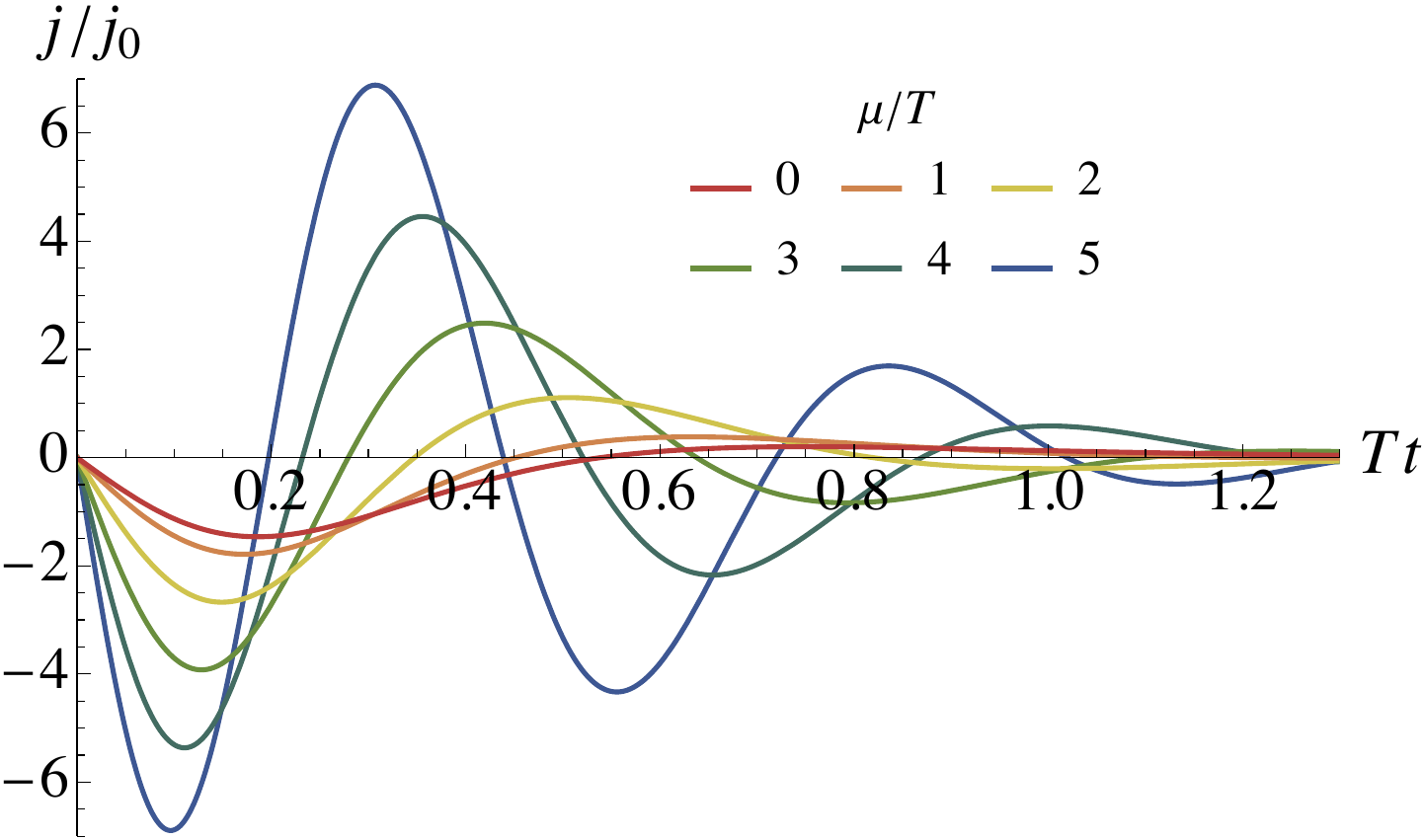}
\caption{\label{fig:pulse}
$j(t)=\hat\b\cdot\j(t)$ at $t>0$ resulting from Eq.~(\ref{eq:pulse}) in units of $j_0\equiv e^3\phi_5bT/(2\pi^2)$ as a function of $Tt$ for $\mu/T=0,1,\dots,5$, where larger $\mu/T$ corresponds to a curve with larger amplitude.
Its integral $\int_{t>0}dt\,j(t)=-e^3\phi_5b/(6\pi^2)$ is independent of $T$ and $\mu$.}
\end{figure}

Extending the latter case further, a uniform but time-dependent magnetic field $\B(t)$ drives the current density in the form of
\begin{align}
\tilde\j(\omega) = \frac{e^3}{2\pi^2}\phi_5\tilde\B(\omega)
\left[1 + \int_0^\infty\!d\eps\,N'_+(\eps)
\frac{4\eps^2(4\eps^2-3\omega^{+2})}{3(4\eps^2-\omega^{+2})^2}\right],
\end{align}
besides the anomalous Hall current due to an induced electric field $\E(x)=-\dot\B(t)\times\r/2$.
The total electric charge transported through a unit area is thus provided by
\begin{align}
\int_{-\infty}^\infty\!dt\,\j(t)
= \frac{e^3}{3\pi^2}\phi_5\int_{-\infty}^\infty\!dt\,\B(t),
\end{align}
being universal in the sense of its independence from the temperature and the chemical potential.
In particular, the time evolution of the current density driven by pulsed $\B(t)=\b\,\delta(t)$,
\begin{align}\label{eq:pulse}
\j(t) &= \frac{e^3}{2\pi^2}\phi_5\b
\left[\delta(t) + \int_0^\infty\!d\eps\,N'_+(\eps)\right. \notag\\
&\quad \times \!\left.\frac{4\eps}{3}
\left[\sin(2\eps t)+\eps t\cos(2\eps t)\right]\theta(t)\right],
\end{align}
is shown in Fig.~\ref{fig:pulse} for various choices of $\mu/T$.
Here, the temporal Friedel oscillation is found to develop with the growing amplitude by increasing $\mu/T$.

\section{Chiral magnetic instability}\label{sec:instability}
A complementary approach to describe low-energy electromagnetic responses of Weyl semimetals is the chiral kinetic theory~\cite{Stephanov:2012,Son:2012,Son:2013b}, where the nonequilibrium distribution function $f_\p(x)$ of particles obeys
\begin{align}
\frac{\d f_\p(x)}{\d t} + \dot\r(x)\cdot\grad_{\!\r}f_\p(x)
+ \dot\p(x)\cdot\grad_{\!\p}f_\p(x) = -\frac{\delta f_\p(x)}{\tau}
\end{align}
in the relaxation-time approximation with
\begin{subequations}
\begin{align}
\dot\r(x) &= \frac{\v_\p(x) + e\E_\p(x)\times\bm\Omega_\p
+ e\left[\v_\p(x)\cdot\bm\Omega_\p\right]\B(x)}{1 + e\B(x)\cdot\bm\Omega_\p}, \\
\dot\p(x) &= \frac{e\E_\p(x) + e\v_\p(x)\times\B(x)
+ e^2\left[\E_\p(x)\cdot\B(x)\right]\bm\Omega_\p}{1 + e\B(x)\cdot\bm\Omega_\p}.
\end{align}
\end{subequations}
Here, $\bm\Omega_\p=\chi\hat\p/(2\p^2)+O(B,E)$ is the Berry curvature with $\chi=\pm$ being the chirality, $\eps_\p(x)=v|\p|[1-e\B(x)\cdot\bm\Omega_\p]+O(B^2,BE)$ is the quasiparticle energy, $\v_\p(x)=\grad_{\!\p}\eps_\p(x)$ is the quasiparticle velocity, $\E_\p(x)=\E(x)-\grad_{\!\r}\eps_\p(x)/e$ is the effective electric field, and $\tau>0$ is a relaxation time.
The distribution function of antiparticles obeys the same kinetic equation but under the replacement of $e\to-e$ and $\bm\Omega_\p\to-\bm\Omega_\p$.
The charge and current densities are provided by
\begin{subequations}\label{eq:densities}
\begin{align}
\rho(x) &= \delta\rho(x) + \sum_\chi\sum_{p,a}
e\int_\p\left[1 + e\B(x)\cdot\bm\Omega_\p\right]f_\p(x), \\
\j(x) &= \delta\j(x) + \sum_\chi\sum_{p,a}
e\int_\p\left[1 + e\B(x)\cdot\bm\Omega_\p\right]\dot\r(x)f_\p(x) \notag\\
&\quad + \sum_\chi\sum_{p,a}e\grad_{\!\r}\times\int_\p\,\bm\Omega_\p\eps_\p(x)f_\p(x),
\end{align}
\end{subequations}
where $\sum_{p,a}$ denotes the summation over particles and antiparticles and the Chern-Simons current in Eq.~(\ref{eq:chern-simons}) must be added to ensure the local conservation of electric charge~\cite{Gorbar:2017a}.

We now determine the deviation from the local equilibrium distribution function $\delta f_\p(x)=f_\p(x)-n_T[\eps_\p(x)\mp\mu_\chi]$ at the linear order in $\E(x)$ and $\B(x)$, where the upper (lower) sign is for particles (antiparticles) with $\mu_\chi=\mu+\chi\mu_5$~\cite{footnote}.
By linearizing the kinetic equation and then performing the Fourier transformation, we obtain
\begin{align}
\delta\tilde f_\p(k)
= \frac{e\tilde\E(k)\cdot\hat\p+i\omega|\p|[e\tilde\B(k)\cdot\bm\Omega_\p]}
{i\left(\omega_\tau-v\k\cdot\hat\p\right)}\frac{dn_T(v|\p|\mp\mu_\chi)}{d|\p|},
\end{align}
which substituted into Eq.~(\ref{eq:densities}) leads to
\begin{align}
& \tilde\rho(k) = \frac{e^3}{2\pi^2}\A_5\cdot\tilde\B(k) \notag\\
&\quad + \frac{3\epsilon_0\Omega_e^2}{2}\left[g_1(\omega_\tau,|\k|)
- g_3(\omega_\tau,|\k|)\right]\frac{\k\cdot\tilde\E(k)}{i\omega_\tau^2}
\end{align}
besides the equilibrium charge density $\rho_0=e\,(\pi^2T^2+\mu^2+3\mu_5^2)\,\mu/(3\pi^2v^3)$ and
\begin{align}\label{eq:kinetic}
&\tilde\j(k) = -\frac{e^3}{2\pi^2}\A_5\times\tilde\E(k) \notag\\
&\quad + \frac{e^3\phi_5+e^2\mu_5}{2\pi^2}\tilde\B(k)
+ \frac{e^2\mu_5}{2\pi^2}g_1(\omega_\tau,|\k|)
\frac\omega{\omega_\tau}\tilde\B(k) \notag\\
&\quad + \frac{e^2v}{8\pi^2}\left[\frac23 + g_1(\omega_\tau,|\k|)
\frac\omega{\omega_\tau}\right]i\k\times\tilde\B(k) \notag\\
&\quad + \frac{3\epsilon_0\Omega_e^2}{2}
\left[g_1(\omega_\tau,|\k|)\frac{\tilde\E(k)}{i\omega_\tau}
- g_3(\omega_\tau,|\k|)\frac{\hat\k\,[\hat\k\cdot\tilde\E(k)]}{i\omega_\tau}\right].
\end{align}
Here, $\omega_\tau\equiv\omega+i/\tau$, the plasma frequency $\Omega_e^2\equiv e^2(\pi^2T^2/3+\mu^2+\mu_5^2)/(3\pi^2\epsilon_0v)$, and
\begin{align}
g_n(\omega,k) \equiv \left[\frac{n\omega^2-(vk)^2}{2(vk)^2}
\ln\!\left(\frac{\omega+vk}{\omega-vk}\right) - \frac{n\omega}{vk}\right]\frac\omega{vk}
\end{align}
are introduced and the latter's limit reads $\lim_{k\to0}g_n(\omega,k)=(n-3)/3+O[(vk/\omega)^2]$ for later use.
We note that our minimal relaxation-time approximation violates the local conservation of electric charge and it can be cured by incorporating the local shift of chemical potential~\cite{Satow:2014,Stephanov:2015}.
However, we confirmed that such modification does not affect our analysis below at long wavelength because the correction is adding a term proportional to $\k\,[\k\cdot\tilde\E(k)]/\tau$ to the current density as well as multiplying the last term of the charge density by $\omega_\tau/\omega$.

The first three terms on the right-hand side of Eq.~(\ref{eq:kinetic}) for $\mu_5=-e\phi_5$ are at the linear order in $A_5^\lambda$.
In order to show their relation with Eq.~(\ref{eq:response}), let us recall that particles and antiparticles typically have energies of $\eps\sim T,\mu$ and the chiral kinetic theory assumes external fields varying sufficiently slow over space-time compared to $T,\mu$.
Therefore, $|\omega|,v|\k|\ll\eps$ holds, where Eq.~(\ref{eq:response}) is indeed reduced to the first three terms in Eq.~(\ref{eq:kinetic}) for $1/\tau=0^+$ because of
\begin{align}
\lim_{\eps\to\infty}f_\eps(\omega,k) = 1 + g_1(\omega,k),
\end{align}
which ensures the mutual consistency between our two approaches.
In particular, the third term in Eq.~(\ref{eq:kinetic}) provides the dynamical chiral magnetic current.
In its derivation in terms of the chiral kinetic theory, we observed one half arising from the nonequilibrium distribution function driven by a dynamical magnetic field $\sim\B(x)$ and the other half arising from that driven by an induced electric field $\sim\grad\times\E(x)$, which constitutes physical contents of our dynamical chiral magnetic current.
On the other hand, the second term in Eq.~(\ref{eq:kinetic}) consists of the Chern-Simons current and the chiral magnetic effect in Eq.~(\ref{eq:CME}) arising from the equilibrium distribution function, which exactly cancel out in equilibrium~\cite{Gorbar:2017a}.

Although the chiral kinetic theory requires $|\omega_\tau|,v|\k|\ll T$ or $\mu$ for its validity, the current density in Eq.~(\ref{eq:kinetic}) is applicable to nonequilibrium states with $\mu_5\neq-e\phi_5$ and also incorporates dissipation within the relaxation-time approximation.
Its physical consequence can be manifested in the uniform limit,
\begin{align}\label{eq:dissipation}
\lim_{|\k|\to0}\tilde\j(k) &= -\sigma_H\hat\A_5\times\tilde\E(k)
+ \left[\sigma_M - \frac{e^2\mu_5}{3\pi^2}
\frac{\omega\tau}{\omega\tau+i}\right]\tilde\B(k) \notag\\
&\quad + \sigma_E\frac{i}{\omega\tau+i}\tilde\E(k),
\end{align}
where the electric conductivity turns into the Drude form with finite $\sigma_E=\epsilon_0\Omega_e^2\tau$ at zero frequency.
Importantly, the dynamical chiral magnetic current in Eq.~(\ref{eq:uniform}) is found to be valid for $\omega\tau\gg1$ but extinguished by the dissipation for $\omega\tau\ll1$, so that the chiral magnetic conductivity remains nonzero at $\sigma_M=(e^3\phi_5+e^2\mu_5)/(2\pi^2)$ only when Weyl node populations deviate from equilibrium.
In contrast, the anomalous Hall conductivity is provided by $\sigma_H=e^3|\A_5|/(2\pi^2)$ at any frequency independent of the relaxation time.

Whereas the electromagnetic fields are so far considered to be external, their dynamics can be studied according to Maxwell's equation,
\begin{align}
\grad\times\B(x) - \mu_0\epsilon_0\frac{\d\E(x)}{\d t} = \mu_0\j(x),
\end{align}
coupled with Eq.~(\ref{eq:dissipation}) in the temporal gauge for the sake of convenience.
Because the left-hand side is at the second order in derivatives with respect to $\A(x)$ but the right-hand side is at the first order, the latter must vanish by itself at $O(k)$.
By solving the characteristic equation for a plane wave $\A(x)=\A_{\omega,\k}\,e^{-i\omega t+i\k\cdot\r}$ with $k_\parallel=\k\cdot\hat\A_5$ and $k_\perp^2=\k^2-k_\parallel^2$, we find two nontrivial solutions
\begin{align}\label{eq:dispersion}
\omega = \frac{\sigma_M\sigma_Hk_\parallel \pm i\sigma_M\sqrt{\sigma_E^2k_\parallel^2
+ (\sigma_H^2+\sigma_E^2)k_\perp^2}}{\sigma_H^2+\sigma_E^2} + O(k^2)
\end{align}
at low frequency and long wavelength $|\omega|,v|\k|\ll1/\tau\ll T$ or $\mu$.
Remarkably, the resulting dispersion relations are determined only by the three dc conductivities and one of them has a positive imaginary part.

\begin{figure}[t]
\includegraphics[width=0.8\columnwidth]{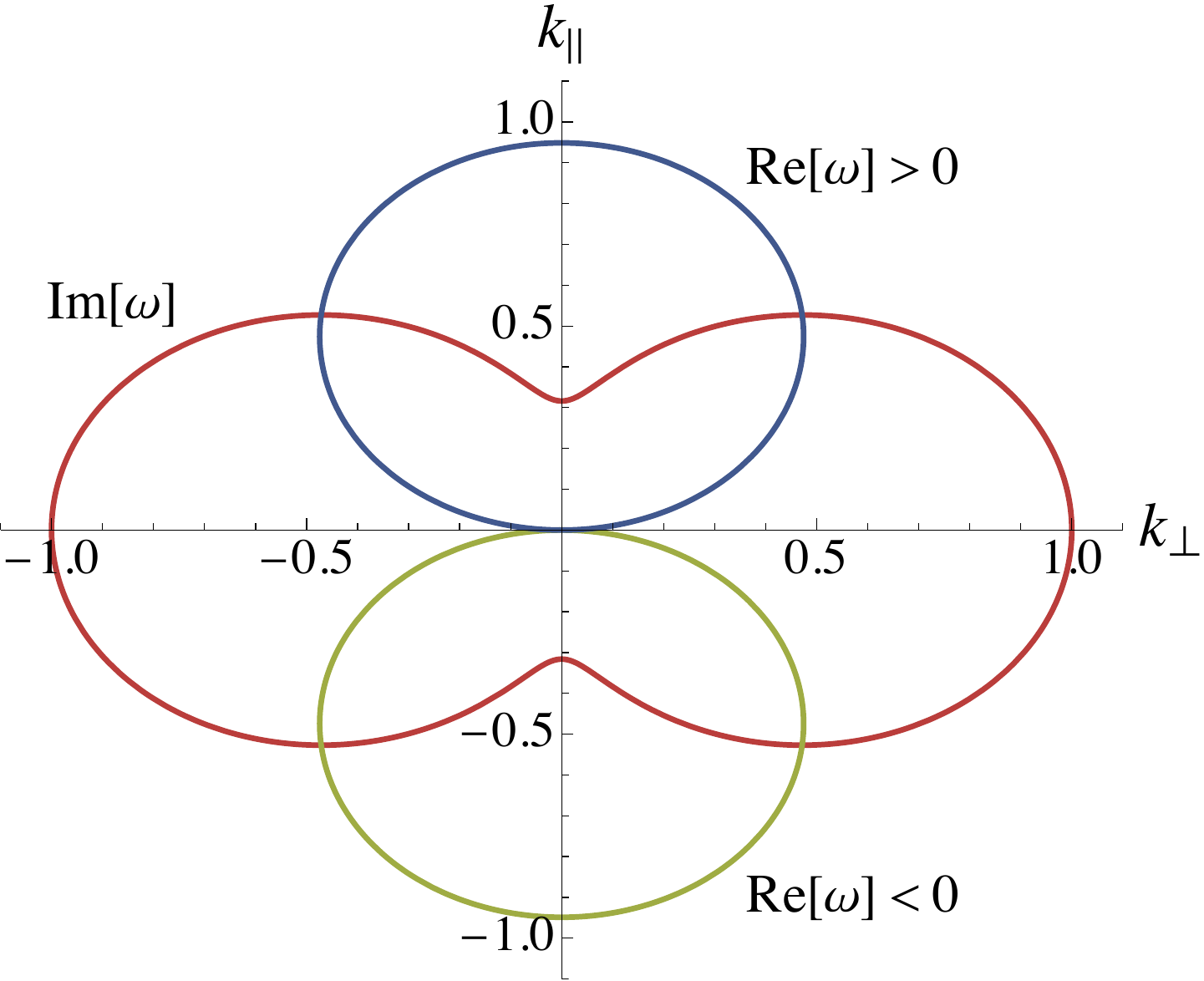}
\caption{\label{fig:anisotropy}
Directional dependence of real and imaginary parts of $\omega(k_\perp,k_\parallel)$ resulting from Eq.~(\ref{eq:dispersion}) for $\sigma_M>0$ and $\sigma_H=3\sigma_E$.
Their magnitudes in units of $\omega_0\equiv\sigma_M|\k|/\sqrt{\sigma_H^2+\sigma_E^2}$ are represented by distances from the origin in the direction of $(k_\perp,k_\parallel)$.}
\end{figure}

The positive imaginary part in frequency leads to the exponential growth of electromagnetic fields over time, which shows notable anisotropy depending on the direction of $\k$ relative to $\A_5$ as exemplified in Fig.~\ref{fig:anisotropy}.
The imaginary part is maximal for $k_\parallel=0$, where the real part vanishes and unstable modes are thus nonpropagating with null Poynting vector, which is also the case for any $\k$ if $\A_5=\0$.
They become propagating waves for $k_\parallel\neq0$ and, in particular, prove to be circularly polarized for $k_\perp=0$, where the imaginary part is minimal but nonzero.
The sign of the real part also depends on $\sigma_Mk_\parallel$ so that the propagation of unstable modes is toward $\pm\hat\k$ for $\pm\sigma_Mk_\parallel>0$ and thus oriented to the direction of $\sigma_M\A_5$, for which both chiral magnetic and anomalous Hall conductivities are essential.
Such instability of the system is analogous to the chiral plasma instability~\cite{Akamatsu:2013}, although its physical origin is different from ours in some details because unstable modes therein were predicted at $\omega\sim\k^2$ in the absence of dissipation.
The resulting exponential growth of electromagnetic fields over time due to $\sigma_M\neq0$ is in turn considered to reduce Weyl node populations toward equilibrium $\mu_5\to-e\phi_5$~\cite{Akamatsu:2013,Manuel:2015,Hirono:2015}, so as to gradually attenuate our chiral magnetic instability.

\section{Summary}
In summary, we studied electromagnetic linear responses of Weyl semimetals based on two of their low-energy effective descriptions, which are the effective field theory and the chiral kinetic theory.
Both approaches consistently showed that a dynamical magnetic field is capable of driving an electric current along its direction, although the static chiral magnetic effect is absent under regularizations.
In particular, the total transported charge is universal for a uniform field in the sense of its independence from temperature and chemical potential, but it was found to be extinguished by dissipation incorporated as a relaxation time.

Furthermore, we studied dispersion relations of collective excitations coupled with Maxwell electromagnetic fields at low frequency and long wavelength.
They are determined only by electric, chiral magnetic, and anomalous Hall conductivities, which predict unstable modes when Weyl node populations deviate from equilibrium.
If wave vectors are perpendicular to the direction of Weyl node separation, unstable modes are nonpropagating and otherwise become propagating waves.
Their propagation is not radial but oriented to or opposite to the direction of Weyl node separation depending on the sign of the chiral magnetic conductivity.
Therefore, whereas the instability is caused by the chiral magnetic effect, it is the anomalous Hall effect that causes the propagation.
Such anisotropic generation of electromagnetic waves from Weyl semimetals is possibly observable by driving Weyl node populations out of equilibrium and its detailed study will be reported elsewhere~\cite{Nishida:preprint}.

\acknowledgments
This work was supported by JSPS KAKENHI Grants No.\ JP18H05405 and No.\ JP21K03384.
One of the authors (T.A.) was also supported by RIKEN Junior Research Associate Program.

\end{document}